\documentclass[11pt,preprint,showpacs]{revtex4}
\usepackage{graphicx}
\usepackage{dcolumn}
\usepackage{bm}
\usepackage{amsmath}
\usepackage{amssymb}
\usepackage{amsfonts}

\begin{document}

\title{Treatment of the proton-proton  Coulomb force in
  proton-deuteron breakup Faddeev calculations}

\author{H.~Wita{\l}a}
\affiliation{M. Smoluchowski Institute of Physics, Jagiellonian
University,
                    PL-30059 Krak\'ow, Poland}

\author{R.~Skibi\'nski}
\affiliation{M. Smoluchowski Institute of Physics, Jagiellonian
University,
                    PL-30059 Krak\'ow, Poland}

\author{J.~Golak}
\affiliation{M. Smoluchowski Institute of Physics, Jagiellonian
University,
                    PL-30059 Krak\'ow, Poland}

\author{W.\ Gl\"ockle}
\affiliation{Institut f\"ur theoretische Physik II,
Ruhr-Universit\"at Bochum, D-44780 Bochum, Germany}

\date{\today}

\begin{abstract}
We extend our approach to incorporate the proton-proton (pp)
Coulomb force into the three-nucleon (3N) Faddeev calculations from
elastic proton-deuteron (pd) scattering to the
 breakup process. 
The main new ingredient is a 3-dimensional screened pp  Coulomb
t-matrix obtained by a numerical solution of the 3-dimensional
Lippmann-Schwinger (LS) equation. We demonstrate numerically 
  that the  proton-deuteron (pd) breakup 
observables can be determined from the resulting on-shell 
3N amplitudes increasing the screening radius. However, contrary to the pd 
elastic scattering, the screening limit
exists only after renormalisation of the pp t-matrices.
\end{abstract}


\maketitle \setcounter{page}{1}

\section{Introduction}
 \label{intro}

The  long-range nature of the Coulomb force 
prevents the application of the standard techniques developed for
short-range interactions   in  the analysis of nuclear
reactions involving two protons. 
 One proposal to avoid the
difficulties  including the Coulomb force is to use a screened
Coulomb interaction and to reach the pure Coulomb limit through
application of a renormalisation procedure~\cite{Alt78,Alt96,Alt94,Alt2002}.

For  elastic pd
scattering first calculations, with modern nuclear forces and the
Coulomb force included, have been achieved in a variational
hyperspherical harmonic approach~\cite{kievski}. Only recently the
inclusion of the Coulomb force was undertaken also for the pd
breakup reaction~\cite{delt2005br}. In~\cite{delt2005br}, contrary
to~\cite{kievski} where the exact Coulomb force in coordinate
representation has been used directly, a screened pp Coulomb force
has been applied in momentum space and in a partial wave basis. In
order to get the final predictions which can be compared to the
data, the limit to the unscreened situation has been performed
numerically applying a renormalization to the resulting
 3N on-shell amplitudes~\cite{delt2005el,delt2005br}. This allowed for the
first time to analyze high-precision pd breakup data and provided
a significant improvement of data description in cases where the
Coulomb force plays an important role~\cite{stephan}.

Recently we developed a novel approach to include the pp Coulomb
force into the momentum space 3N Faddeev calculations~\cite{elascoul}. 
 It is based on a standard
 formulation for short range forces and relies on the screening  of the
long-range Coulomb interaction. In order to avoid  all uncertainties
connected with the application of the partial wave expansion, 
 inadequate  when working
with long-range forces, we used directly the 3-dimensional pp screened
Coulomb t-matrix.
We demonstrated the 
feasibility of that approach in case of elastic pd scattering 
using a simple dynamical model for the nuclear part of the interaction. 
 It turned out that the screening limit exists without the need of 
renormalisation not only for pd elastic scattering 
observables but for the elastic pd amplitude itself.

In spite of the substantial progress in the pd breakup treatment 
achieved in~\cite{delt2005el,delt2005br}
 some important
questions remained unanswered. One concerns directly the results
of these calculations for two kinematically complete breakup
geometries: the pp quasi-free-scattering (QFS) configuration, in
which the not detected neutron is at rest in the laboratory system,
and the space-star (SST) geometry, in which all 3 outgoing
nucleons have the same  momenta (magnitudes)  in the plane which in the 3N
c.m. system is
perpendicular  to the incoming nucleon
momentum. The theoretical predictions based on nuclear forces only
show, that the cross sections for QFS and SST are quite stable
against changes of the underlying interactions, including also
three-nucleon forces~\cite{physrep96}. At energies below $\approx 20$~MeV
theory
 overestimates the SST pd cross sections by $\approx 10\%$,
and overestimates the pp QFS cross sections by $\approx 20\%$,
respectively~\cite{physrep96,delt2005br}. With increasing energy  the
theoretical cross sections
come close to the data~\cite{physrep96}, which  indicates that the pp Coulomb
force is very probably responsible for these low energy
discrepancies. However, the Coulomb force effects found
in~\cite{delt2005br} are practically negligible for the pd SST
configuration and only slightly improve the description of the pp
QFS data~\cite{exp1,exp2,exp3,exp4,exp5}.

This inability to understand the pp QFS and pd SST cross sections
 motivated us to reconsider the  inclusion of the Coulomb force
into momentum space Faddeev calculations. One main concern in such
type of calculations is the application of a partial wave decomposition
to the long-ranged  Coulomb force. Even when screening is applied
it  seems reasonable to treat from the beginning the screened pp
Coulomb t-matrix without partial wave decomposition because the required
limit
 of vanishing screening leads necessarily to  a drastic increase
of the number of partial wave states involved. 
 As an example we provide numbers for the exponential screening 
 $e^{-(\frac {r} {R})^n}$. Taking the screening radius $R=20$~fm and 
$n=4$ requires all $l \le l_{max}=10$ partial wave states to reproduce the 
$3$-dimensional pp screened Coulomb t-matrix at $E_p^{lab}=13$~MeV. 
Increasing the screening radius to $R=120$~fm 
requires  $l_{max}\approx 50$ which is 
 a big numerical challenge. Even more, that would lead to an explosion of the 
number of 3N partial waves required for convergence.  
 Another problem concerns the treatment of the pp Coulomb
interaction in its proper coordinate, which is the relative
proton-proton distance, throughout the  calculations. Any
deviation from this restriction can cause effects which are
difficult to estimate. Both problems were accounted for in our formulation 
presented in~\cite{elascoul} and applied there to pd elastic scattering. 

In the present paper we extend that  approach to 
 the pd breakup. Again 
 we apply directly the 3-dimensional screened pp Coulomb
t-matrix without relying on a partial wave decomposition. 
 In \cite{elascoul} 
we demonstrated 
that the physical pd elastic  scattering amplitude can be obtained from
the off-shell solutions of the Faddeev equation and has a well
defined screening limit. In
contrast to elastic scattering, where the amplitude itself 
 does not require renormalisation, in case of the pd breakup the
on-shell solutions of the Faddeev equation are required. They 
 demand renormalisation in the screening limit which can be achieved 
 through renormalisation of the pp t-matrices.

In section \ref{form} for the convenience of the reader  we shortly 
present  the main points of the formalism outlined in details 
in \cite{elascoul}. 
 The screening limit for pd breakup 
is discussed in section \ref{limit} and the results shown
in section \ref{results}. The summary is given in section
\ref{sumary}.

\section{Faddeev equations with screened pp Coulomb force}
\label{form}

We use the Faddeev equation in the form~\cite{physrep96} 
\begin{eqnarray}
T| \Phi> =  t P | \Phi> +  t P G_0 T| \Phi>
\label{13a}
\end{eqnarray} 
where $ P $ is defined in terms of transposition operators,
 $ P =P_{12} P_{23} + P_{13} P_{23} $, $G_0$ is the free 3N propagator,
 $ |\Phi>$ the initial state composed of a deuteron state and a momentum
 eigenstate of the proton.
Knowing $T| \Phi>$ the breakup as well as the elastic pd scattering amplitudes
 can be gained by quadratures in the standard manner~\cite{physrep96}.

We use our standard momentum space partial wave basis $|pq\tilde {\alpha}>$
\begin{eqnarray}
|p q \tilde {\alpha}> \equiv  = |pq(ls)j(\lambda
\frac {1} {2})I (jI)J (t \frac {1} {2})T>\label{51} 
\end{eqnarray}
and  distinguish between the partial wave states $|pq\alpha>$ with
total 2N angular momentum $j$ below some value $j_{max}$: $j \le
j_{max}$,
 in which the nuclear, $V_N$, as well as the pp screened Coulomb
interaction, $V_c^R$
 (in isospin $t=1$ states only), are acting,  and the
states $|pq\beta>$ with $j > j_{max}$, for which only $V_c^R$ is
acting in the pp subsystem. The states $|pq\alpha>$ and
$|pq\beta>$ form a complete system of states
\begin{equation}
\int {p^2 dpq^2 dq} \sum\limits_{\tilde \alpha}  {\left| {pq{\tilde \alpha} }
\right\rangle } \left\langle {pq{\tilde \alpha} } \right| = 
\int {p^2 dpq^2 dq} (\sum\limits_\alpha  {\left| {pq\alpha }
\right\rangle } \left\langle {pq\alpha } \right| +
\sum\limits_\beta  {\left| {pq\beta } \right\rangle } \left\langle
{pq\beta } \right|) = {\rm I} \\~ . 
\label{52}
\end{equation}
Projecting Eq.(\ref{13a}) for $ T| \Phi >$ on the $|pq\alpha>$ and
$|pq\beta>$ states one gets the following system of coupled
integral equations
\begin{eqnarray}
 \left\langle {pq\alpha } \right|T\left| {\Phi } \right\rangle  &=&
 \left\langle {pq\alpha } \right|t_{N + c}^R P\left| {\Phi } \right\rangle
\cr
 &+& \left\langle {pq\alpha } \right|t_{N + c}^R PG_0 \sum\limits_{\alpha '}
 {\int {p'^2 dp'q'^2 dq'\left| {p'q'\alpha '} \right\rangle \left\langle
 {p'q'\alpha '} \right|} } T\left| {\Phi } \right\rangle \cr
 &+& \left\langle {pq\alpha } \right|t_{N + c}^R PG_0 \sum\limits_{\beta '}
 {\int {p'^2 dp'q'^2 dq'\left| {p'q'\beta '} \right\rangle \left\langle
 {p'q'\beta '} \right|} } T\left| {\Phi } \right\rangle  \label{53} \\
 \left\langle {pq\beta } \right|T\left| {\Phi } \right\rangle  &=&
 \left\langle {pq\beta } \right|t_c^R P\left| {\Phi } \right\rangle \cr
  &+& \left\langle {pq\beta } \right|t_c^R PG_0 \sum\limits_{\alpha '}
  {\int {p'^2 dp'q'^2 dq'\left| {p'q'\alpha '} \right\rangle
  \left\langle {p'q'\alpha '} \right|} } T\left| {\Phi } \right\rangle \cr
  &+& \left\langle {pq\beta } \right|t_c^R PG_0 \sum\limits_{\beta '}
  {\int {p'^2 dp'q'^2 dq'\left| {p'q'\beta '} \right\rangle
  \left\langle {p'q'\beta '} \right|} } T\left| {\Phi } \right\rangle
 \label{54}
\end{eqnarray}
where $t_{N+c}^R$ and $t_c^R$ are t-matrices generated by
the interactions $V_N+V_c^R$ and $V_c^R$, respectively.
 Namely for states $|\alpha >$ with two-nucleon subsystem isospin 
$t=1$ the corresponding t-matrix element 
$<p\alpha |t_{N+c}^R(E-\frac {3} {4m}q^2)|p'\alpha'>$ is a linear 
combination of the pp, $t_{pp+c}^R$, and the neutron-proton (np), 
$t_{np}$, $t=1$ t-matrices, which are generated by the interactions 
$V_{pp}^{strong}+V_c^R$ and $V_{np}^{strong}$, 
respectively. The 
coefficients of that combination depend on the total isospin $T$ and
$T'$  
of states $|\alpha >$ and $|\alpha' >$~\cite{elascoul,wit91}:
\begin{eqnarray}
<t=1T=\frac{1} {2} |t_{N+c}^R|t'=1T'=\frac{1}
{2}> 
& =&  \frac {1} {3} t_{np} + \frac {2} {3} t_{pp+c}^{R} \cr
<t=1T=\frac{3} {2} |t_{N+c}^R|t'=1T'=\frac{3}
{2}> 
& =&  \frac {2} {3} t_{np} + \frac {1} {3} t_{pp+c}^{R} \cr
<t=1T=\frac{1} {2} |t_{N+c}^R|t'=1T'=\frac{3}
{2}> 
& =&  \frac {\sqrt{2}} {3}( t_{np} - t_{pp+c}^{R}) \cr
<t=1T=\frac{3} {2} |t_{N+c}^R|t'=1T'=\frac{1}
{2}> 
& =&  \frac {\sqrt{2}} {3}( t_{np} - t_{pp+c}^{R})    ~.
\label{8}
\end{eqnarray}
For isospin $t=0$, in which case 
$T=T'=\frac {1} {2}$:
\begin{eqnarray}
<t=0T=\frac{1} {2} |t_{N+c}^R|t'=0T'=\frac{1}
{2}> 
& =& t_{np} ~.
\label{8a}
\end{eqnarray}
In case of $t_c^R$ only the screened pp Coulomb force $V_c^R$ is acting.

The third term  on the right hand side of (\ref{54}) is
proportional to $<pq\beta|t_c^RPG_0|p'q'\beta'><p'q'\beta'|t_c^ R$. A direct
calculation of its isospin part shows that independently from the
value of the total isospin $T$ it vanishes~\cite{elascoul}. 

Inserting $<pq\beta|T|\Phi>$ from (\ref{54}) into (\ref{53}) one
gets
\begin{eqnarray}
 \left\langle {pq\alpha } \right|T\left| {\Phi } \right\rangle  &=&
 \left\langle {pq\alpha } \right|t_{N + c}^R P\left| {\Phi } \right\rangle
 + \left\langle {pq\alpha } \right|t_{N + c}^R PG_0 t_c^R P\left| {\Phi }
\right\rangle \cr
  &-& \left\langle {pq\alpha } \right|t_{N + c}^R PG_0 \sum\limits_{\alpha
'}
  {\int {p'^2 dp'q'^2 dq'\left| {p'q'\alpha '} \right\rangle
  \left\langle {p'q'\alpha '} \right|} } t_c^R P\left| {\Phi } \right\rangle
\cr
  &+& \left\langle {pq\alpha } \right|t_{N + c}^R PG_0 \sum\limits_{\alpha
'}
  {\int {p'^2 dp'q'^2 dq'\left| {p'q'\alpha '} \right\rangle
  \left\langle {p'q'\alpha '} \right|} } T\left| {\Phi }
  \right\rangle \cr
  &+& \left\langle {pq\alpha } \right|t_{N + c}^R PG_0 t_c^R PG_0
\sum\limits_{\alpha '}
  {\int {p'^2 dp'q'^2 dq'\left| {p'q'\alpha '} \right\rangle
  \left\langle {p'q'\alpha '} \right|} } T\left| {\Phi } \right\rangle \cr
  &-& \left\langle {pq\alpha } \right|t_{N + c}^R PG_0  \sum\limits_{\alpha
'}
  {\int {p'^2 dp'q'^2 dq'\left| {p'q'\alpha '} \right\rangle
  \left\langle {p'q'\alpha '} \right|} } t_c^R PG_0 \cr
  && \sum\limits_{\alpha'' }
  {\int {p''^2 dp''q''^2 dq''\left| {p''q''\alpha'' } \right\rangle
  \left\langle {p''q''\alpha'' } \right|} } T\left| {\Phi } \right\rangle ~.
  \label{55}
 \end{eqnarray}
This is a coupled set of integral equations in the space of  the states
$|\alpha>$ only, which incorporates the contributions of
the pp Coulomb interaction from all partial wave states up to
infinity. 
 It can be solved by iteration and Pade
summation~\cite{physrep96,elascoul}. 

When compared to our standard treatment without screened Coulomb 
force~\cite{physrep96}  there are two new leading terms
$<pq\alpha|t_{N+c}^RPG_0t_c^RP|\Phi>$  and
-$<pq\alpha|t_{N+c}^RPG_0|\alpha'><\alpha'|t_c^RP|\Phi>$. The
first term must be calculated using directly
the $3$-dimensional screened Coulomb t-matrix $t_c^R$, while the
second term requires partial wave projected screened Coulomb
t-matrix elements in the  $|\alpha>$ channels only. The kernel also contains
two new terms.  The term
$<pq\alpha|t_{N+c}^RPG_0t_c^RPG_0|\alpha'><\alpha'|T|\Phi>$ must again be
calculated with a 3-dimensional screened Coulomb t-matrix while the second
one,
-$<pq\alpha|t_{N+c}^RPG_0|
\alpha'><\alpha'|t_c^RPG_0|\alpha''><\alpha''|T|\Phi>$,  involves only
 the partial wave projected screened Coulomb t-matrix elements in the
$|\alpha>$ channels. The calculation of the new terms with the
partial wave projected Coulomb t-matrices follows our standard
procedure. Namely the  two sub kernels $t_{N+c}^RPG_0$ and
$t_c^RPG_0$ are applied consecutively  on the corresponding state.
The detailed expressions how to calculate the new terms with the  3-dimensional
screened Coulomb t-matrix  are given in Appendix A of Ref.~\cite{elascoul}.

The transition amplitude for 
 breakup $<\Phi_0|U_0|\Phi>$ is given
in terms of $T\left| {\Phi } \right\rangle$ by~\cite{gloeckle83,physrep96}
\begin{eqnarray}
 \left\langle {\Phi _0 } \right|U_0 \left| {\Phi } \right\rangle  &=&
 \left\langle {\Phi _0 } \right|(1 + P)T\left| {\Phi } \right\rangle
\label{56}
 \end{eqnarray}
where $ | \Phi_0> = | \vec p
\vec q m_1 m_2 m_3 \nu_1 \nu_2 \nu_3 > $ is the free state. 
 The permutations acting in momentum-, spin-, and isospin-spaces 
can be applied to the bra-state $< \phi_0| = < \vec p
\vec q m_1 m_2 m_3 \nu_1 \nu_2 \nu_3 | $ changing the sequence of 
nucleons spin and isospin magnetic
quantum numbers $ m_i$ and  $\nu_i$ and leading to well known linear
combinations of the Jacobi momenta $ \vec p,\vec q$. Thus
evaluating (\ref{56}) it is sufficient to regard the general
amplitudes $< \vec p \vec q m_1 m_2 m_3 \nu_1 \nu_2 \nu_3 | 
T\left| {\Phi }\right\rangle \equiv \left\langle {\vec p\vec q~} 
\right|T\left| {\Phi }
\right\rangle $. Using Eq.~(\ref{13a}) and the completness relation (\ref{52})
one gets: 
\begin{eqnarray}
&&\left\langle {\vec p\vec q~} \right|T\left| {\Phi }
\right\rangle  = \left\langle {\vec p\vec q~}
\right|\sum\limits_{\alpha '} {\int {p'^2 dp'q'^2 dq'\left|
{p'q'\alpha '} \right\rangle \left\langle {p'q'\alpha '} \right|}
} T\left| {\Phi } \right\rangle \cr && - \left\langle {\vec
p\vec q~} \right|\sum\limits_{\alpha '} {\int {p'^2 dp'q'^2
dq'\left| {p'q'\alpha '} \right\rangle \left\langle {p'q'\alpha '}
\right|} } t_c^R P\left| {\Phi } \right\rangle \cr &&-
\left\langle {\vec p\vec q~} \right|\sum\limits_{\alpha '} {\int
{p'^2 dp'q'^2 dq'\left| {p'q'\alpha '} \right\rangle \left\langle
{p'q'\alpha '} \right|} } t_c^R PG_0
 \sum\limits_{\alpha
''} {\int {p''^2 dp''q''^2 dq''\left| {p''q''\alpha ''}
\right\rangle \left\langle {p''q''\alpha ''} \right|} } T\left|
{\Phi } \right\rangle \cr && + \left\langle {\vec p\vec q~}
\right|t_c^R P\left| {\Phi } \right\rangle  + \left\langle
{\vec p\vec q~} \right|t_c^R PG_0 \sum\limits_{\alpha '} {\int
{p'^2 dp'q'^2 dq'\left| {p'q'\alpha '} \right\rangle \left\langle
{p'q'\alpha '} \right|} } T\left| {\Phi } \right\rangle ~.
\label{69}
\end{eqnarray}

It follows, that in addition to the amplitudes $<pq\alpha|T|\Phi >$ also the
partial wave projected amplitudes $<pq\alpha|t_c^RP|\Phi >$ and
$<pq\alpha|t_c^RPG_0|\alpha'><\alpha'|T|\Phi>$ are required. The
expressions for the contributions of these three terms to the
transition amplitude for the breakup reaction are
given in Appendix B of Ref.~\cite{elascoul}.

The last two terms in (\ref{69}) again must be calculated using
directly $3$-dimensional screened Coulomb t-matrices. In Appendix C  
of Ref.~\cite{elascoul} the expression for 
$\left\langle {\vec p\vec q~} \right|t_c^R
P\left| {\Phi } \right\rangle$ 
 and in Appendix D of Ref.~\cite{elascoul} 
 the last matrix element $<\vec p \vec q~ |t_c^RPG_0|\alpha
'><\alpha '|T|\Phi>$ are given.

\section{The screening limit}
\label{limit}

The set of coupled Faddeev equations (\ref{55}) 
 is well defined for any finite screening radius. It is
an exact set assuming that the strong NN t-matrix can be
neglected beyond a certain $ j_{max}$, which is   justified. Further
the pp screened Coulomb force is taken  into account to infinite
order in the partial wave decomposition in form of the 3-dimensional 
screened Coulomb t-matrix $ t_{pp}^{cR}$. The
important challenge is to control the screening  limit for the
physical pd breakup  amplitude (\ref{56}). In case of elastic
scattering we provided in Ref.~\cite{elascoul} analytical arguments and showed 
 numerically that the physical elastic pd scattering amplitude itself 
 has a well 
defined screening limit and does not require renormalisation. 
This can be traced back to the fact that to get the elastic pd scattering 
amplitude it is sufficient to solve the Faddeev 
equations (\ref{55}) for off-shell values of the Jacobi momenta 
\begin{eqnarray}
    \frac {p^2} {m} + \frac{3}{4m} q^2 \ne  E ~.
\label{73}
\end{eqnarray}
The off-shell Faddeev amplitudes 
 $\left\langle {pq\alpha } \right|T\left| {\Phi } \right\rangle$ of 
Eq.(\ref{55}) are determined by off-shell nucleon-nucleon t-matrix elements 
$t(p,p';E-\frac {3} {4m}q^2)$, which have a well defined screening limit 
(see the following discussion and examples). 

Contrary to  pd elastic scattering the physical breakup 
amplitude (\ref{56}) corresponds to the on-shell values of Jacobi momenta 
\begin{eqnarray}
    \frac {p^2} {m} + \frac{3}{4m} q^2 =  E \equiv \frac{3}{4m} q_{max}^2 ~.
\label{74}
\end{eqnarray}
That means that the physical pd breakup amplitude (\ref{69}) requires 
on-shell Faddeev amplitudes 
$\left\langle {p_0q\alpha } \right|T\left| {\Phi } \right\rangle$ together with
 the four, also on-shell, additional terms in (\ref{55}), with 
$p_0=\sqrt {\frac {3} {4} (q_{max}^2-q^2)}$. 
 The on-shell Faddeev amplitudes can be obtained from the off-shell solutions 
$\left\langle {pq\alpha } \right|T\left| {\Phi } \right\rangle$ using 
(\ref{55}):
\begin{eqnarray}
 \left\langle {p_0q\alpha } \right|T\left| {\Phi } \right\rangle  &=&
 \left\langle {p_0q\alpha } \right|t_{N + c}^R P\left| {\Phi } \right\rangle
 + \left\langle {p_0q\alpha } \right|t_{N + c}^R PG_0 t_c^R P\left| {\Phi }
\right\rangle \cr
  &-& \left\langle {p_0q\alpha } \right|t_{N + c}^R PG_0 \sum\limits_{\alpha
'}
  {\int {p'^2 dp'q'^2 dq'\left| {p'q'\alpha '} \right\rangle
  \left\langle {p'q'\alpha '} \right|} } t_c^R P\left| {\Phi } \right\rangle
\cr
  &+& \left\langle {p_0q\alpha } \right|t_{N + c}^R PG_0 \sum\limits_{\alpha
'}
  {\int {p'^2 dp'q'^2 dq'\left| {p'q'\alpha '} \right\rangle
  \left\langle {p'q'\alpha '} \right|} } T\left| {\Phi }
  \right\rangle \cr
  &+& \left\langle {p_0q\alpha } \right|t_{N + c}^R PG_0 t_c^R PG_0
\sum\limits_{\alpha '}
  {\int {p'^2 dp'q'^2 dq'\left| {p'q'\alpha '} \right\rangle
  \left\langle {p'q'\alpha '} \right|} } T\left| {\Phi } \right\rangle \cr
  &-& \left\langle {p_0q\alpha } \right|t_{N + c}^R PG_0  \sum\limits_{\alpha
'}
  {\int {p'^2 dp'q'^2 dq'\left| {p'q'\alpha '} \right\rangle
  \left\langle {p'q'\alpha '} \right|} } t_c^R PG_0 \cr
  && \sum\limits_{\alpha'' }
  {\int {p''^2 dp''q''^2 dq''\left| {p''q''\alpha'' } \right\rangle
  \left\langle {p''q''\alpha'' } \right|} } T\left| {\Phi } \right\rangle ~. 
  \label{eq55}
 \end{eqnarray}
These on-shell amplitudes together with additional, also on-shell, 
terms in (\ref{69}) define the physical breakup amplitude (\ref{56}). 
That in consequence requires half-shell t-matrix elements 
$t(p_0,p';\frac {p_0^2} {m})$ which are of 3 types:  
the partial wave projected pure screened Coulomb $t_c^R$ generated by 
$V_{c}^R$, the 
 partial wave projected  $t_{N+c}^R$ generated by
 $V_{strong}+V_{c}^R$, 
and the 3-dimensional screened Coulomb matrix elements. 

It is well known \cite{Alt78,ford1964,ford1966}  that in the screening limit 
 $ R \rightarrow \infty$ such half-shell t-matrices  
acquire an infinitely oscillating phase factor $ e^ {i \Phi_R(p)}$,
 where $\Phi_R(p)$ depends on the type of the screening. 
 For the exponential screening its form  depends
on two parameters, the screening radius $R$ and the power $n$:
\begin{equation}
V_c^R(r) = \frac{\alpha} {r} e^{-{(\frac {r} {R})}^n} ~.
\label{eq.2}
\end{equation}
At a given value $n$ the pure Coulomb potential results for $R \rightarrow
\infty$. As has been shown in~\cite{kamada05} based
on~\cite{taylor1,taylor2},  the related phase $ \Phi_R(p)$ is given as
\begin{eqnarray}
\Phi_R(p) = -\eta [ ln(2pR) - {\epsilon}/n] 
\label{eq.8}
\end{eqnarray}
where  $\epsilon=0.5772\dots$ is the 
Euler number and $\eta = \frac{ m_p \alpha}{2
p} $ the Sommerfeld
parameter.

Contrary to the half-shell, the off-shell t-matrix elements do not 
acquire such an 
oscillating phase and their screening limit is well defined. 

\begin{figure}
\includegraphics[scale=0.8]{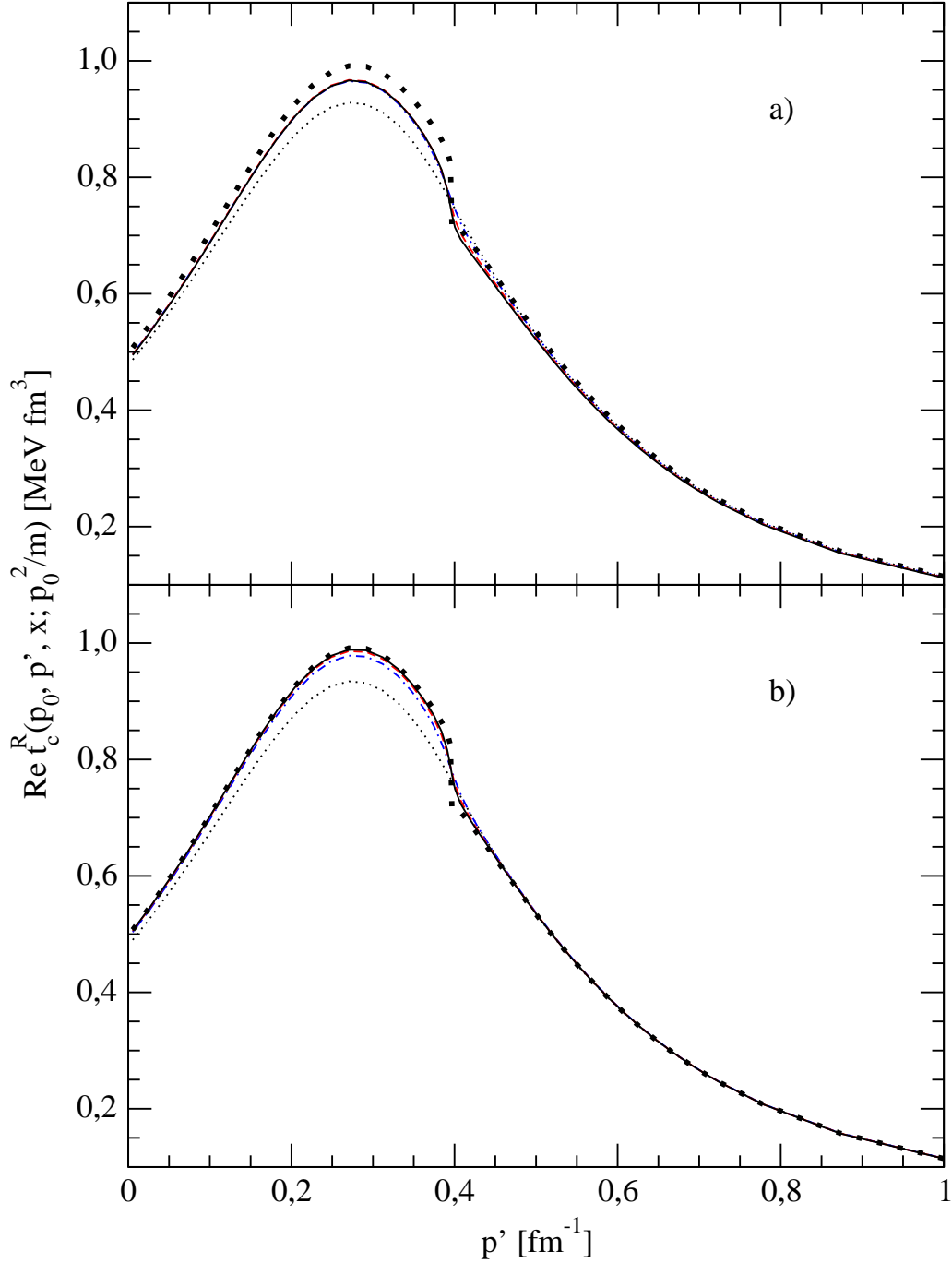}
\caption{(color online) 
The real part of the nonrenormalized (a) and renormalized (b) 
 3-dimensional half-shell screened  Coulomb 
pp t-matrix $t_c^R(p_0,p',x;\frac {p_0^2} {m})$. The lines
correspond to the exponential screening with $n=1$ and 
different screening radia: 
 $R=20$~fm (black dotted line), 
 $R=60$~fm (blue dashed-dotted line), 
 $R=120$~fm (red dashed line), 
 $R=180$~fm (black solid line). The 
 Coulomb half-shell 
result of Ref.~\cite{kok81} is given by thick dots. 
The momentum $p_0=0.396$~fm$^{-1}$ and
$x=0.706$.}
\label{fig1}
\end{figure}

\begin{figure}
\includegraphics[scale=0.8]{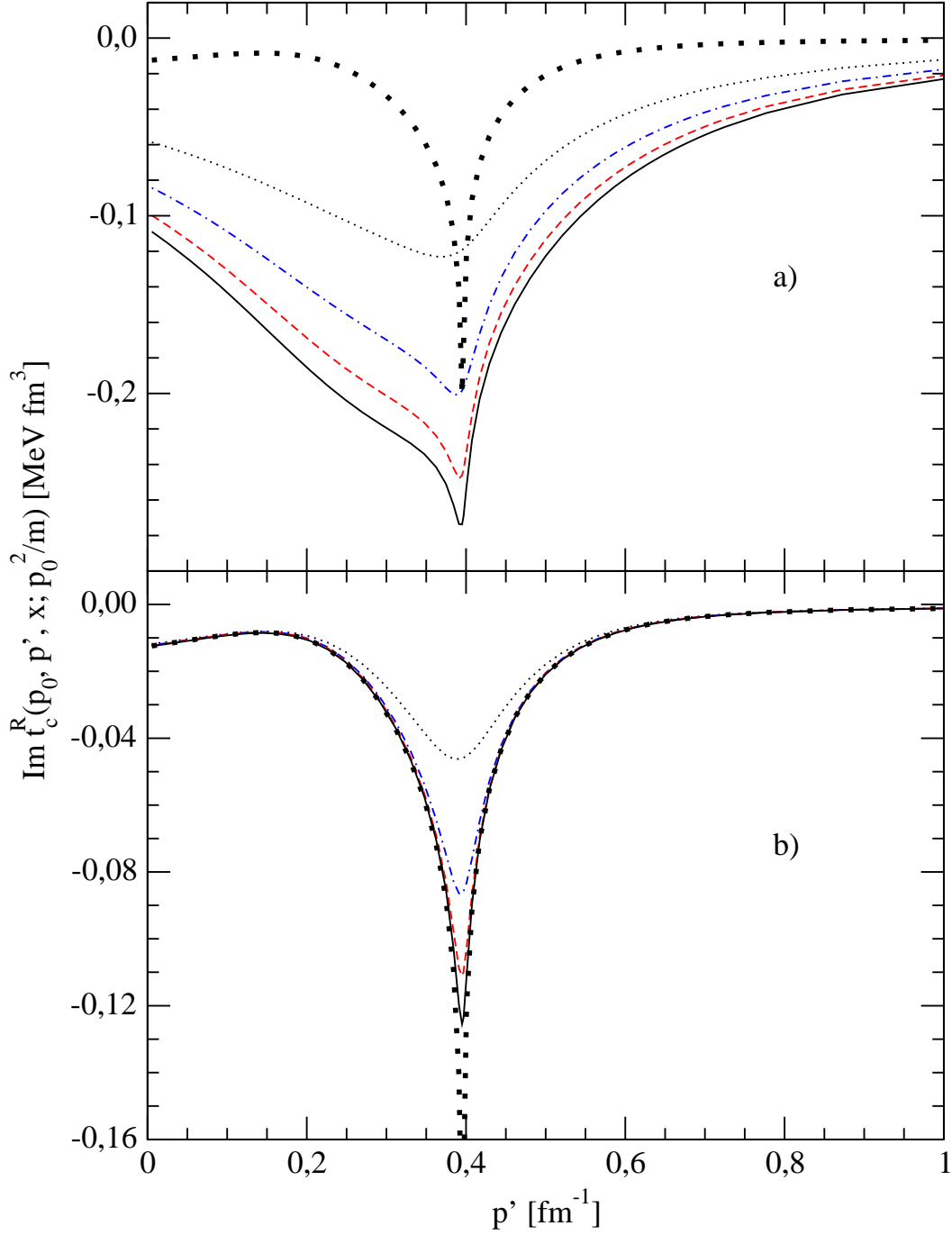}
\caption{(color online) 
The imaginary part of the nonrenormalized (a) and renormalized (b) 
 3-dimensional half-shell screened  Coulomb 
pp t-matrix $t_c^R(p_0,p',x;\frac {p_0^2} {m})$.
 For the description of the 
lines and values of $p_0$ and $x$ see Fig.~\ref{fig1}.}
\label{fig2}
\end{figure}

\begin{figure}
\includegraphics[scale=0.9]{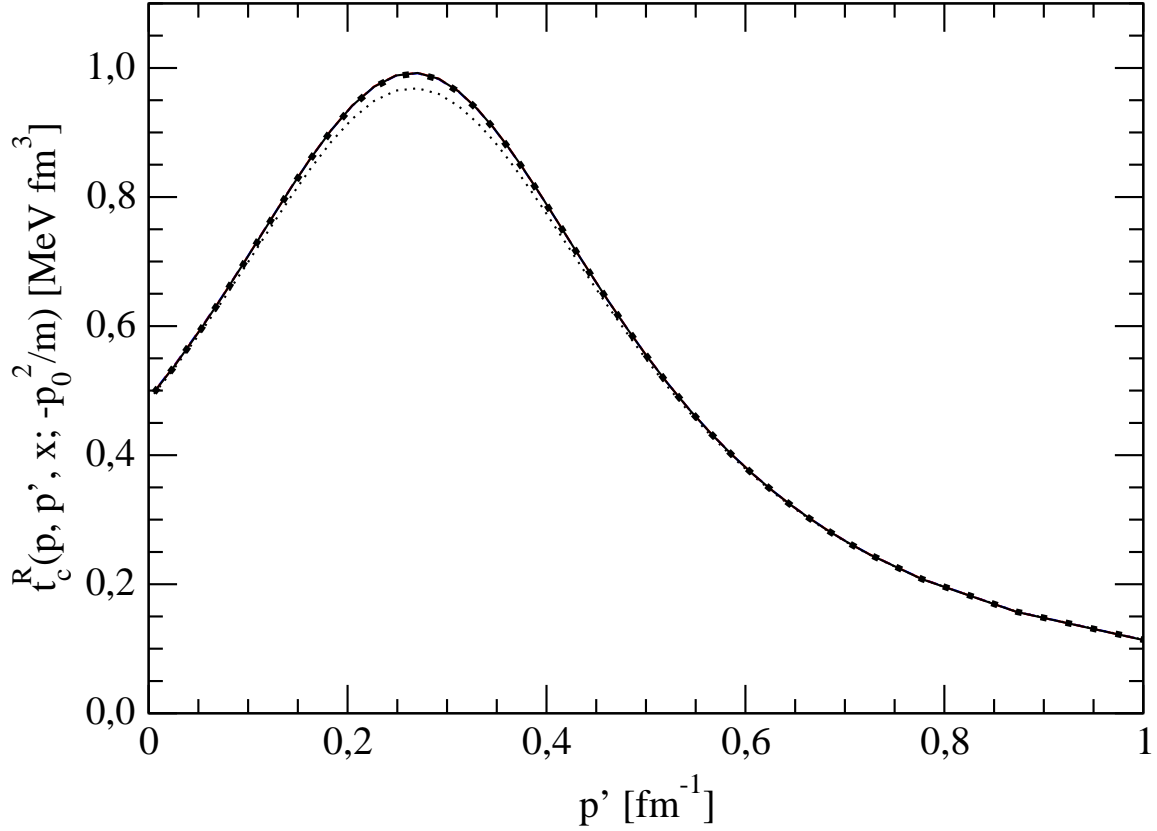}
\caption{(color online) 
The  3-dimensional  off-shell screened  Coulomb 
pp t-matrix $t_c^R(p,p',x;-\frac {p_0^2} {m})$. The lines
correspond to the exponential screening with $n=1$ and 
different screening radia: 
 $R=20$~fm (black dotted line), 
 $R=60$~fm (blue dashed-dotted line), 
 $R=120$~fm (red dashed line), 
 $R=180$~fm (black solid line). The 
 Coulomb off-shell 
 result of Ref.~\cite{chen72,kok1980} is given by thick dots. 
The momentum $p_0=0.396$~fm$^{-1}$, $p=0.375$~fm$^{-1}$ and
$x=0.706$. 
}
 \label{fig3}
\end{figure}

In Figs.~\ref{fig1}-\ref{fig3} we demonstrate that 
behavior for the 3-dimensional half-shell 
screened Coulomb pp t-matrix~\cite{Witala08}. 
Increasing the screening radius R changes 
drastically the imaginary part of the t-matrix (Fig.~\ref{fig2}a). The real 
part is more stable but does not approach the pure Coulomb limit  
(Fig.~\ref{fig1}a). Renormalizing by the phase factor $ e^ {-i \Phi_R(p)}$ of 
Eq.~(\ref{eq.8}) provides a well defined limit to the pure Coulomb 
half-shell result of Ref.~\cite{kok81} (Fig.~\ref{fig1}b and \ref{fig2}b). 

For the  3-dimensional off-shell screened Coulomb pp t-matrix 
the pure Coulomb screening limit  of Ref.~\cite{chen72,kok1980} is 
achieved without any renormalisation factor for screening radia 
$R > 20$~fm (Fig.~\ref{fig3}).

\begin{figure}
\includegraphics[scale=0.9]{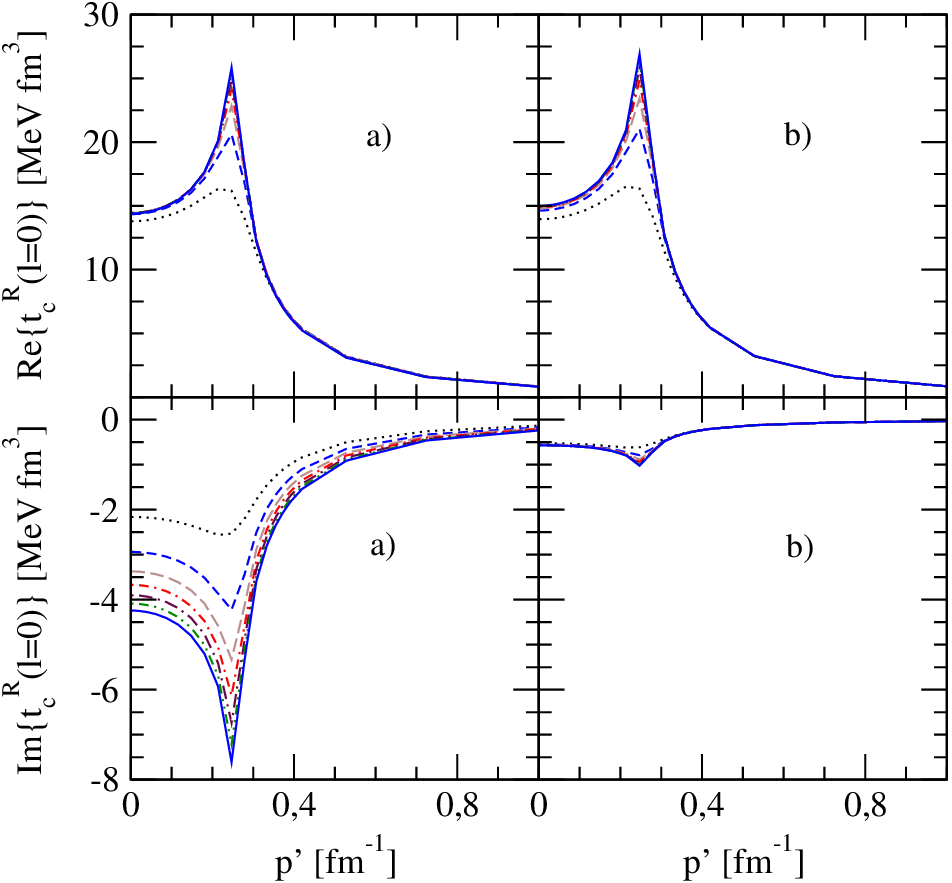}
\caption{(color online) 
The real (upper panels) and imaginary (lower panels) parts of the 
 nonrenormalized (a) and renormalized (b) $l=0$ half-shell  
screened  Coulomb 
pp t-matrix $t_c^R(p_0,p',\frac {p_0^2} {m})$. The lines
correspond to the exponential screening with $n=1$ and 
different screening radia: 
 $R=20$~fm (black dotted line), 
 $R=40$~fm (blue short-dashed line), 
 $R=60$~fm (brown long-dashed line), 
 $R=80$~fm (red short-dashed-dotted line), 
 $R=100$~fm (maroon long-dashed-dotted line), 
 $R=120$~fm (green short-dashed-double-dotted line), 
 $R=140$~fm (blue solid line). 
 The momentum $p_0=0.26$~fm$^{-1}$.
}
\label{fig4}
\end{figure}

\begin{figure}
\includegraphics[scale=0.9]{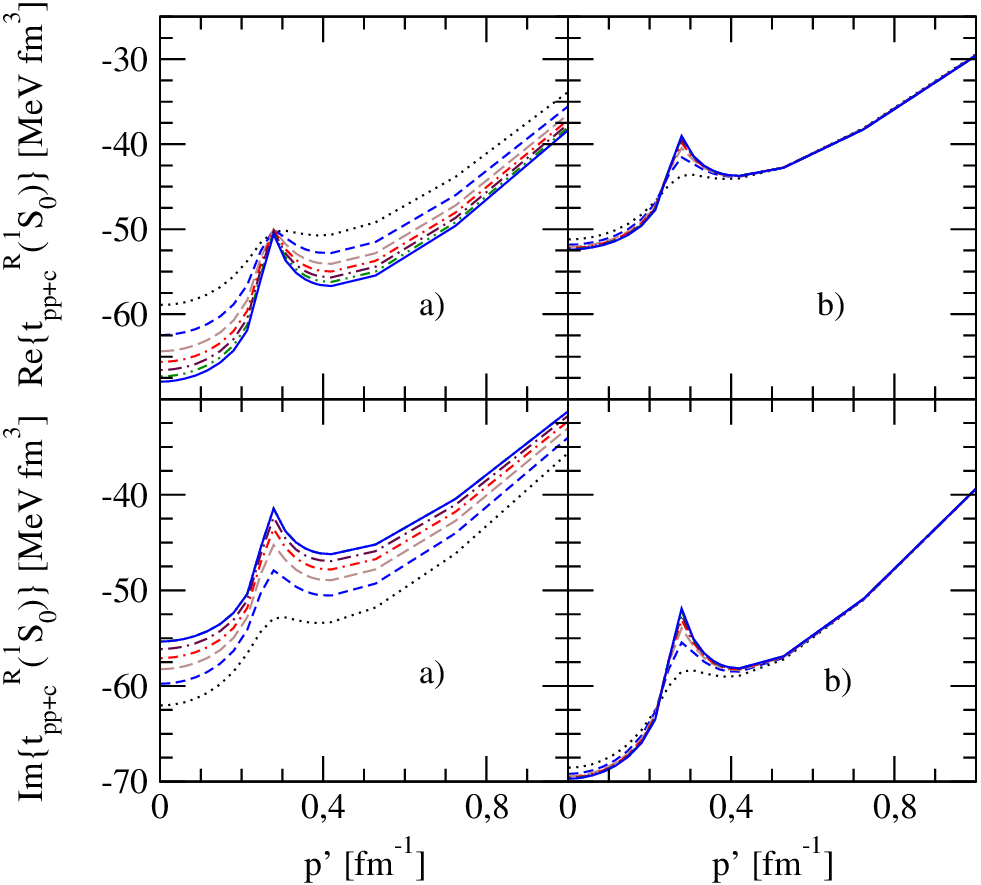}
\caption{(color online) 
The real (upper panels) and imaginary (lower panels) parts of the 
 nonrenormalized (a) and renormalized (b) $^1S_0$ half-shell  
  pp t-matrix $t_{pp+c}^R(p_0,p',\frac {p_0^2} {m})$. The lines
correspond to the exponential screening with $n=1$ and 
different screening radia: 
 $R=20$~fm (black dotted line), 
 $R=40$~fm (blue short-dashed line), 
 $R=60$~fm (brown long-dashed line), 
 $R=80$~fm (red short-dashed-dotted line), 
 $R=100$~fm (maroon long-dashed-dotted line), 
 $R=120$~fm (green short-dashed-double-dotted line), 
 $R=140$~fm (blue solid line). 
 The momentum $p_0=0.26$~fm$^{-1}$.
}
\label{fig5}
\end{figure}

Analogous behavior for the  partial wave decomposed $l=0$ 
half-shell screened Coulomb $t_c^R$ and 
 the $^1S_0$   $t_{pp+c}^R$ t-matrices  is shown in 
 Figs.~\ref{fig4} and \ref{fig5}, respectively. While the 
imaginary part again 
exhibits drastic changes when the screening radius increases 
(Fig.~\ref{fig4}a and \ref{fig5}a), removing 
the phase factor $ e^ {-i \Phi_R(p)}$ (renormalisation) 
provides a well defined limit 
 for the screening radia $R > 40$~fm (Fig.~\ref{fig4}b and \ref{fig5}b). 
It is seen that in case when the screened Coulomb potential 
is combined with the strong force 
also the real part of the half-shell t-matrix undergoes strong changes with 
increased screening (Fig.~\ref{fig5}a).

\begin{figure}
\includegraphics[scale=0.9]{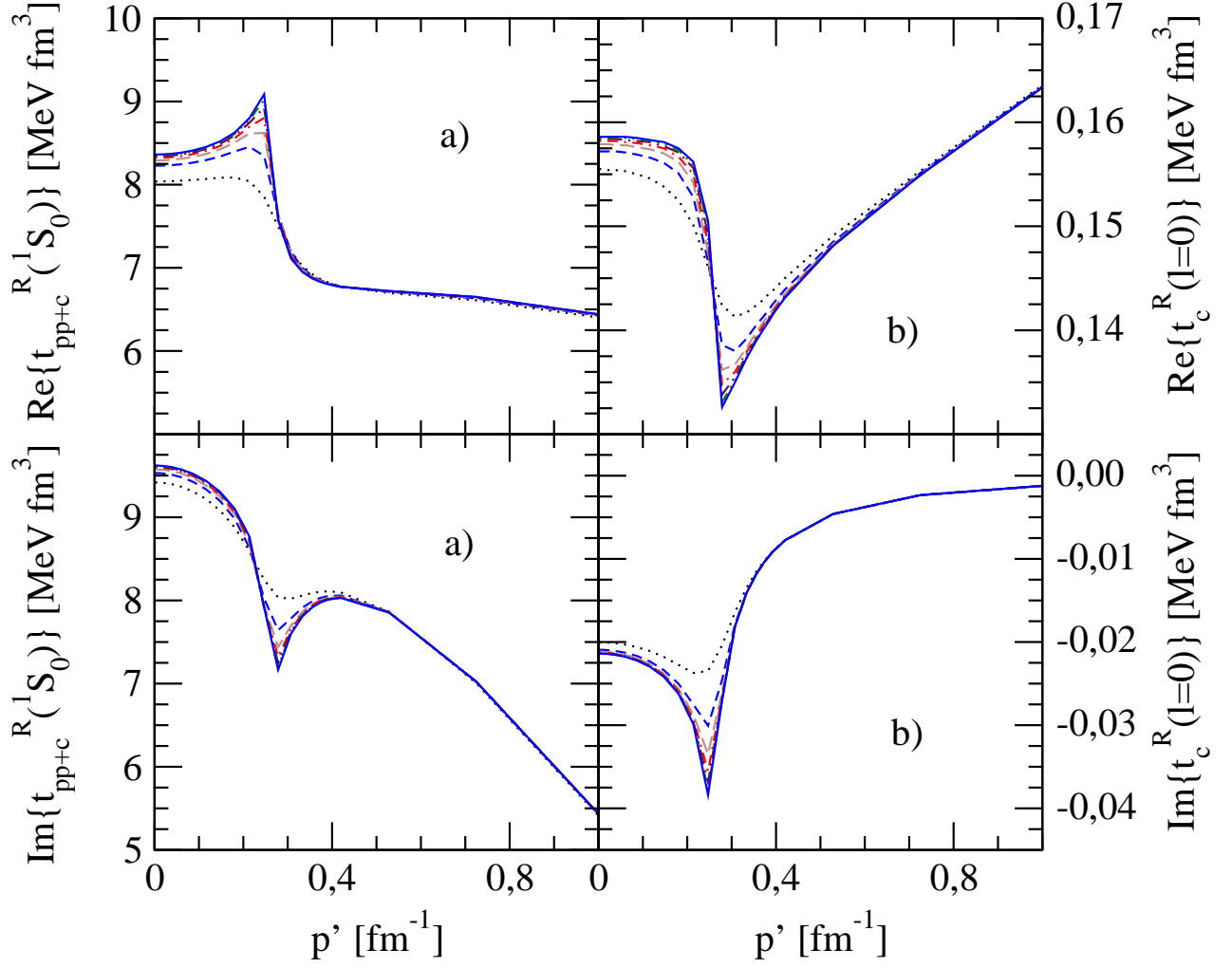}
\caption{(color online) 
The real (upper panels) and imaginary (lower panels) parts of the 
 $^1S_0$  $t_{pp+c}^R(p,p',\frac {p_0^2} {m})$ (a) 
and  the $l=0$  screened Coulomb $t_c^R(p,p',\frac {p_0^2} {m})$ (b) off-shell  
  t-matrices. 
For the description of the lines see Fig.~\ref{fig4}.
 The momentum $p_0=0.26$~fm$^{-1}$ and $p=2.38$~fm$^{-1}$.
}
\label{fig6}
\end{figure}

For the  partial wave decomposed 
off-shell $l=0$ screened Coulomb $t_c^R$ and the $^1S_0$  $t_{pp+c}^R$ 
  t-matrix elements a well defined 
screening limit is reached without any renormalisation 
(Fig.~\ref{fig6}b and \ref{fig6}a, respectively). 

That oscillatory phase factor appearing in the half-shell
proton-proton 
 t-matrices requires 
 a carefull treatment of (\ref{eq55}) to get 
the screening limit for the  
$\left\langle {p_0q\alpha } \right|T\left| {\Phi } \right\rangle$ 
amplitudes. Namely for the 
states $|\alpha >$ with the two-nucleon subsystem isospin 
$t=1$ the corresponding t-matrix element 
$<p_0\alpha |t_{N+c}^R(\frac {p_0^2} {m})|p'\alpha'>$ is a linear 
combination of the pp and neutron-proton (np) $t=1$ t-matrices,  
 the coefficients of which  depend on the total isospin T and T' 
of the states $|\alpha >$ and $|\alpha' >$ (see discussion after (\ref{54})).
 It follows that to achieve the screening limit one 
needs to renormalize the pp t-matrix $t_{pp+c}^R$ in that combination 
 before performing the action 
of the operators in (\ref{eq55}). The term in that linear combination
 coming with the np t-matrix
$t_{np}$ does not require renormalisation.

\section{Numerical results}
\label{results}

To demonstrate the feasibility of our approach we applied the outlined 
formalism to a simple dynamical model in which the nucleon-nucleon
force was restricted
to act in $^1S_0$ and $^3S_1-^3D_1$ partial waves only  
 and taken from the CD~Bonn potential~\cite{cdbonn}. The
proton-proton Coulomb force was modified by the exponential screening 
 (\ref{eq.2})
 with the screening radius $R$ and $n=1$.

To investigate the screening limit $R \to \infty$ we generated set of
partial-wave decomposed t-matrices, $t_c^R$, based on the screened pp Coulomb
force only  or combined with the strong pp interaction, 
$t_{pp+c}^R$, taking $R=20, 40, 60, 80, 100, 120$ and $140$~fm. With
that dynamical input we solved the set of Faddeev equations (\ref{55}) for 
off-shell values of the Jacobi momenta $p$ and $q$ and for
the total angular momenta of the p-p-n system up to $J \le \frac {15}
{2}$ and both parities. Then the on-shell Faddeev amplitudes 
$\left\langle {p_0q\alpha } \right|T\left| {\Phi } \right\rangle$ 
were gained through (\ref{eq55}).
In this first study we restricted ourselves to
the perturbative approximation for the 3-dimensional screened  Coulomb
t-matrix: $t_c^R = V_c^R$.  Of course in the future studies that
approximation will be avoided and the full solution of the
3-dimensional LS equation for the screened pp Coulomb t-matrix will be
used~\cite{Witala08}. When calculating observables we also omitted  the 
 last term $ \left\langle \Phi_0 |  (1+P)t_c^RPG_0T | {\Phi } \right\rangle $ 
 in (\ref{69}) coming with the 3-dimensional screened Coulomb t-matrix.

\begin{figure}
\includegraphics[scale=0.9]{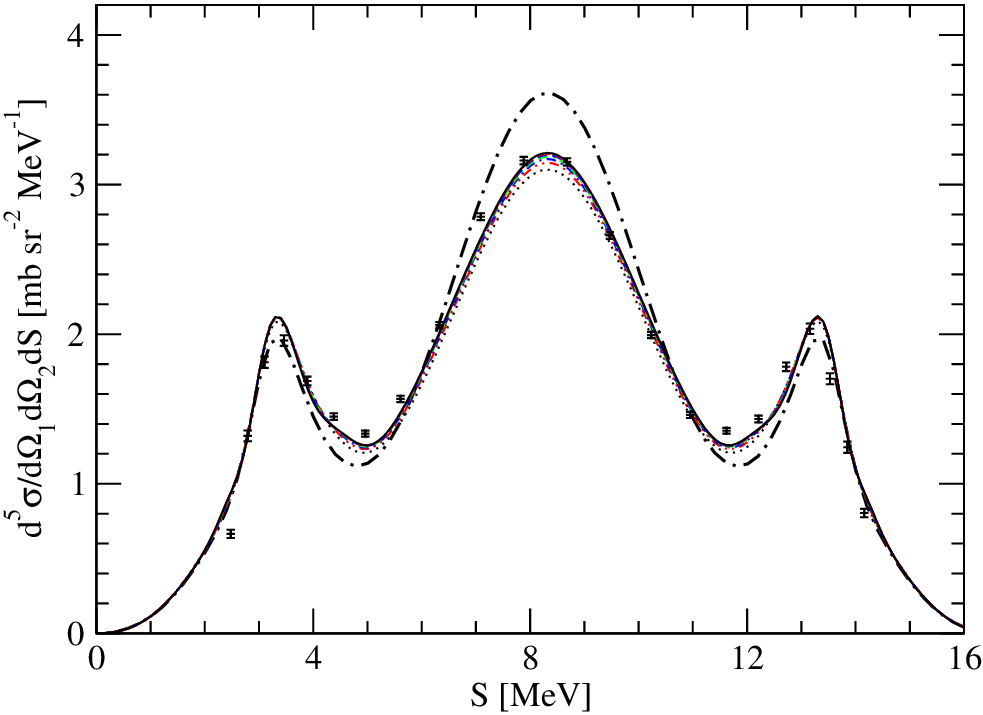}
\caption{(color online) The convergence in the cut-off radius R 
of the  $d(p,p_1p_2)n$ breakup cross section in a kinematically complete 
 QFS configuration with polar angles of the two
outgoing protons $\theta_1 = \theta_2 = 39^o$ and azimuthal angle 
$\phi_{12}=180^o$. 
The incoming proton energy is 
$E_p^{lab}=13$~MeV 
 and theoretical predictions are based on a screened Coulomb
force and the CD~Bonn nucleon-nucleon potential~\cite{cdbonn}
restricted to $^1S_0$ and $^3S_1$-$^3D_1$ partial waves. 
 The screening
 radius  is:  $R=20$~fm (black dotted line), 
 $R=40$~fm (green dashed-double-dotted line), 
 $R=60$~fm (blue dashed-dotted line), 
 $R=80$~fm (red double-dashed-dotted line),
 $R=100$~fm (blue dashed line), 
 $R=120$~fm (red long-dashed line),
 $R=140$~fm (black solid line).  
The black long-dashed-dotted line is the break-up cross section with the 
pp Coulomb interaction switched-off. 
The pluses are $E_p^{lab}=13$~MeV pd breakup 
data of Ref.~\cite{exp2}.} 
\label{fig7}
\end{figure}

\begin{figure}
\includegraphics[scale=0.8]{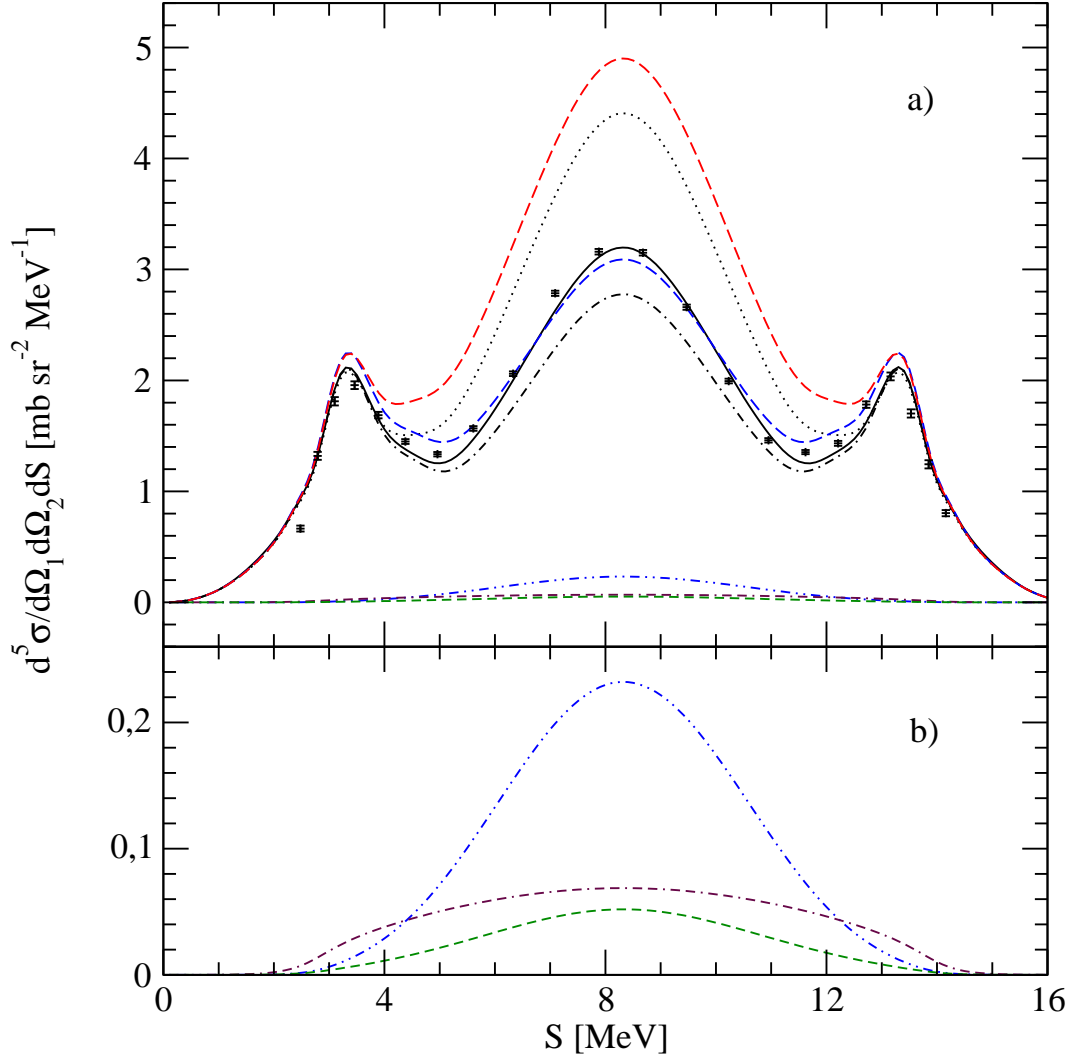}
\caption{(color online) In part a): the contribution of different terms to 
the  cross section of the QFS configuration of Fig.~\ref{fig7}. The
 (black) dashed-dotted line is the contribution of the first term 
 $ \left\langle \Phi_0 |  (1+P)|\alpha><\alpha|T | {\Phi }
\right\rangle $ 
in (\ref{69}) and  
 the (green) short-dashed line  is the
contribution of the fourth term  
$ \left\langle \Phi_0 |  (1+P)t_c^RP | {\Phi } \right\rangle $
 in (\ref{69}) coming with the 3-dimensional screened 
Coulomb t-matrix $t_c^R$ (in the present calculation $t_c^R = V_c^R$). 
The (blue) dashed-double-dotted and (maroon) double-dashed-dotted
lines are contributions of the second   
 $ \left\langle \Phi_0 |  (1+P)|\alpha><\alpha|
t_c^RP | {\Phi } \right\rangle $ and of the third 
 $ \left\langle \Phi_0 |  (1+P)|\alpha><\alpha|
t_c^RPG_0T | {\Phi } \right\rangle $
term in (\ref{69}), respectively, which are calculated 
with partial-wave decomposed 
 screened Coulomb t-matrix.  The
 (black) dotted and (blue) long-dashed lines result from  the 
$ \left\langle \Phi_0 | (1+P)|\alpha><\alpha|
(T - t_c^RP) | {\Phi }  \right\rangle $ 
and 
$ \left\langle \Phi_0 | (1+P)|\alpha><\alpha|
(T - t_c^RPG_0T) | {\Phi }  
\right\rangle $
 amplitudes, respectively. The (red) long-dashed line is the contribution of 
the $ \left\langle \Phi_0 | (1+P)|\alpha><\alpha|
(T - t_c^RP - t_c^RPG_0T) | {\Phi }  
\right\rangle $ amplitude. 
The (black) solid line encompasses all four terms. All results are for
screening radius $R=100$~fm. The part b) of the figure shows
contributions of small terms which are difficult to see on the scale
of part a).
}
\label{fig8}
\end{figure}

\begin{figure}
\includegraphics[scale=0.9]{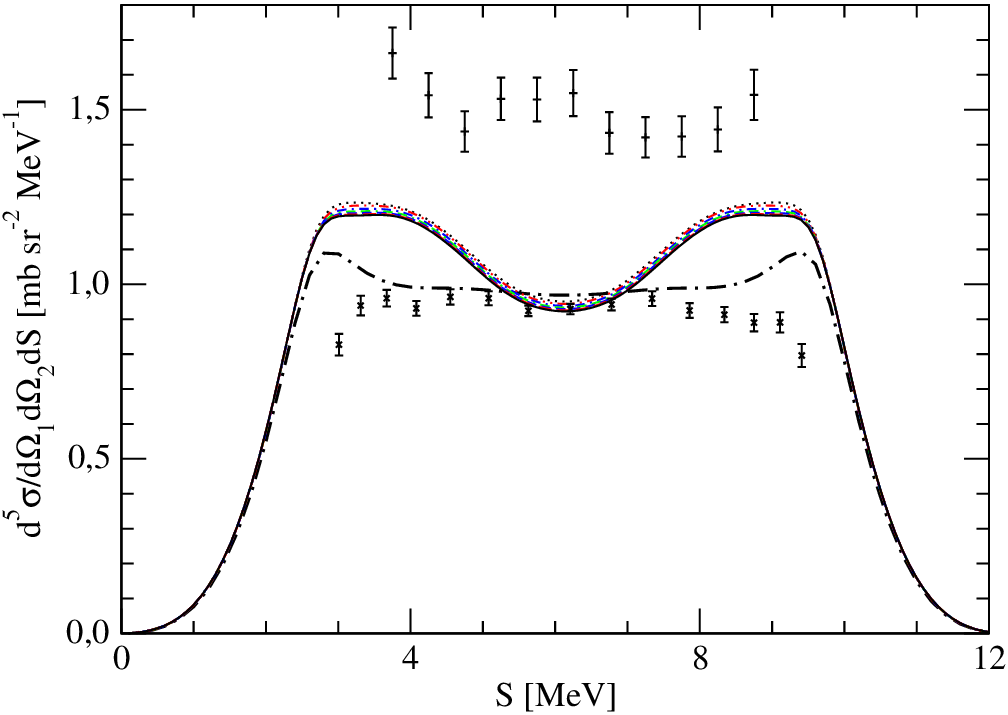}
\caption{(color online) The convergence in the cut-off radius R 
of the  $d(p,p_1p_2)n$ breakup cross section in a kinematically complete 
 SST configuration with polar angles of the two
outgoing protons $\theta_1 = \theta_2 = 50.5^o$ and azimuthal angle 
$\phi_{12}=120^o$. 
The incoming proton energy is 
$E_p^{lab}=13$~MeV 
 and the theoretical predictions are based on the screened Coulomb
force and the CD~Bonn nucleon-nucleon potential~\cite{cdbonn}
restricted to $^1S_0$ and $^3S_1$-$^3D_1$ partial waves. 
 For the description of the lines see Fig.~\ref{fig7}. 
 The x-es are $E_p^{lab}=13$~MeV pd breakup 
data of Ref.~\cite{exp2} and pluses are $E_n^{lab}=13$~MeV 
nd breakup data of Ref.~\cite{tunl}.} 
\label{fig9}
\end{figure}

\begin{figure}
\includegraphics[scale=0.9]{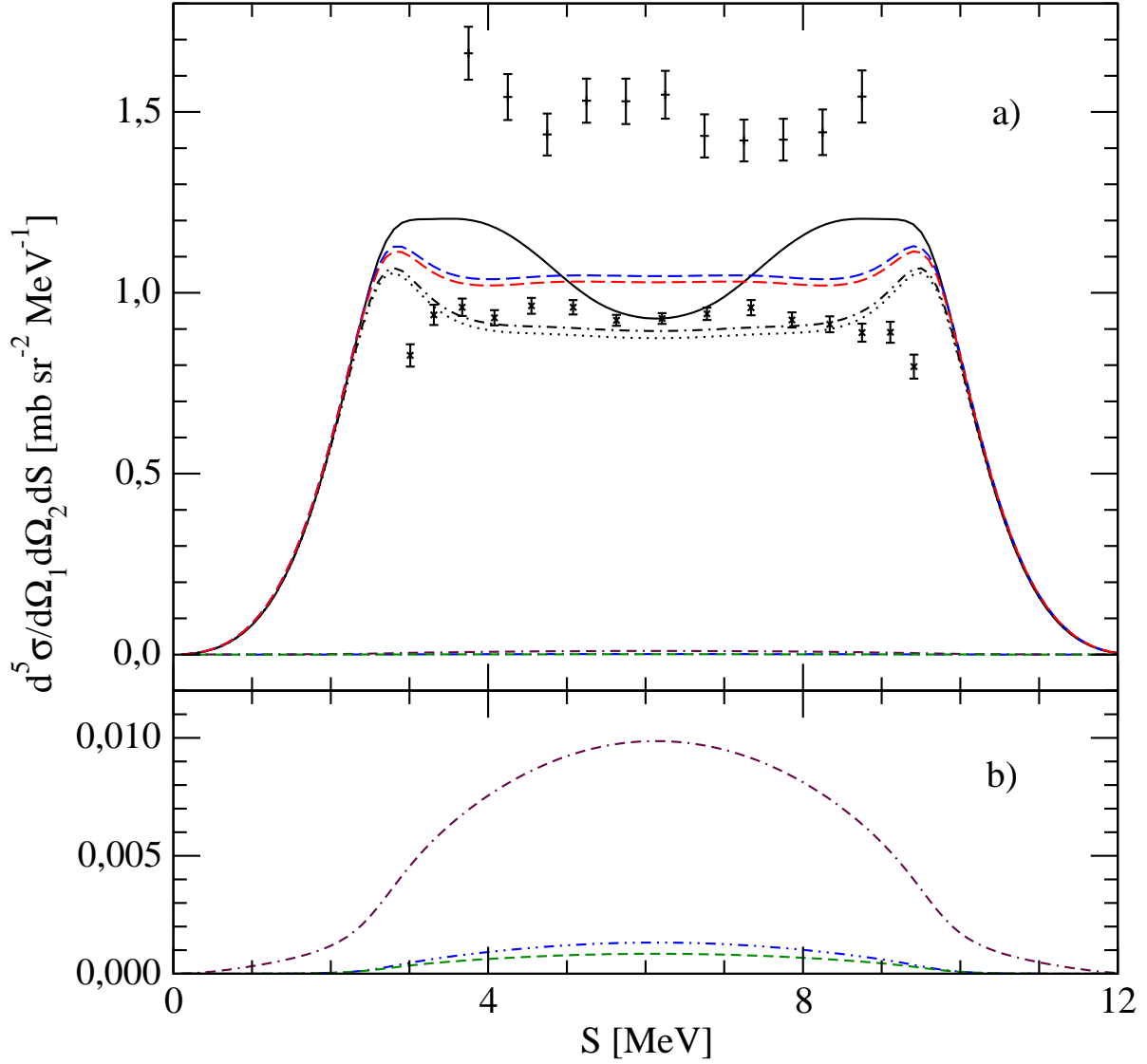}
\caption{(color online) In part a): the contribution of different terms to 
 the  cross section of the SST configuration of Fig.~\ref{fig9}.
 The solid line encompasses all terms. 
 For the 
description of other lines see Fig.~\ref{fig8}. 
All results are for
screening radius $R=100$~fm. The part b) of the figure shows
contributions of small terms which are difficult to see on the scale
of part a).
}
\label{fig10}
\end{figure}

The results for the breakup reaction are shown in
Figs.~\ref{fig7}-\ref{fig10} where the exclusive cross sections 
 $\frac{d^5\sigma} {d\Omega_1 \Omega_2 dS}$
 for the QFS and SST configurations parametrized through the
 arc-length of the kinematical S-curve are presented.  

For the QFS and SST  (see Fig.~\ref{fig7} and \ref{fig9}, respectively) 
the convergence in the
screening radius is 
 achieved at $R=60$~fm. 
 For QFS  the Coulomb force
 decreases the cross section with respect to the nd case and brings the 
theory close to the pd data. For SST the 
Coulomb force also brings theory close 
to the pd data, however, only at S-values close to the space-star condition 
($S \approx 6$~MeV). For S-values further away the theory is far above 
the pd data.

The theoretical prediction for both geometries results through 
  interference of different terms contributing to the breakup amplitude. 
 The importance and 
magnitudes of the contributions coming from different terms in the breakup
amplitude differs for those two geometries (see Fig.~\ref{fig8} and 
Fig.~\ref{fig10}). In both cases the largest is the contribution of the 
 first term in (\ref{69})  
 $ \left\langle \Phi_0 | (1+P)|\alpha><\alpha|T | {\Phi } \right\rangle $ 
(black dashed-dotted line in Figs.~\ref{fig8} and \ref{fig10}). 
 For QFS and SST the cross section resulting from that term is 
 below the pd data and below the full result which encompasses all terms (solid
 line). 
 The magnitudes of three additional terms: 
 $ \left\langle \Phi_0 | (1+P)t_c^RP | {\Phi } \right\rangle $
(the fourth term in (\ref{69}) calculated with the 3-dimensional 
screened Coulomb t-matrix, here 
 approximated by $V_c^R$,  and given by
 the green short-dashed line), the second term in (\ref{69}) 
 $ \left\langle \Phi_0 | (1+P)|\alpha><\alpha|t_c^RP | {\Phi } \right\rangle $
(calculated with the partial-wave projected 
 screened Coulomb t-matrix and given by
the blue dashed-double-dotted line), and the third  term in (\ref{69}) 
  $ \left\langle \Phi_0 | (1+P)|\alpha><\alpha|
t_c^RPG_0T | {\Phi } \right\rangle $
(calculated again with the partial-wave projected 
 screened Coulomb t-matrix and given by
the maroon double-dashed-dotted line), are small. 
 Because they are difficult to see on the scale of Figs.~\ref{fig8}a and 
Fig.~\ref{fig10}a they are again presented in the part b) of these
figures. 
 For both configurations the term with the 3-dimensional t-matrix  
 $ \left\langle \Phi_0 | (1+P)t_c^RP | {\Phi } \right\rangle $ gives
 the smallest contribution. 
 Smallness  of these terms does not mean however, that they are unimportant 
because the interference effects are
nonnegligible and act in different ways for QFS and SST.

For the QFS the second largest contribution comes from
  $ \left\langle \Phi_0 | (1+P)|\alpha><\alpha|t_c^RP | {\Phi } \right\rangle $ 
(blue dashed-double-dotted line) while for SST it comes from  
 $ \left\langle \Phi_0 | (1+P)|\alpha><\alpha|
t_c^RPG_0T | {\Phi } \right\rangle $ 
(maroon double-dashed-dotted line). For QFS and SST taking the
amplitude of that second largest contribution together with 
 $ \left\langle \Phi_0 | (1+P)|\alpha><\alpha|T | {\Phi }
\right\rangle $ 
changes
significantly the cross section. For QFS it is the black dotted line in 
Fig.~\ref{fig8}a resulting from 
 $ \left\langle \Phi_0 | (1+P)|\alpha><\alpha|
(T - t_c^RP) | {\Phi } \right\rangle $ 
while for SST it is blue long-dashed line in Fig.~\ref{fig10}a
resulting from 
 $ \left\langle \Phi_0 | (1+P)|\alpha><\alpha|(T - t_c^RPG_0T) | {\Phi }
 \right\rangle$. 

The third largest contribution provides for both configurations
smaller changes of the cross section and the result when the second
and third largest contributions are included 
 $ \left\langle \Phi_0 | (1+P)|\alpha><\alpha|
(T - t_c^RPG_0T -t_c^RP) | {\Phi } 
\right\rangle $  is given by the red long-dashed line. It is above the
pd data for both geometries. 

Finally, including the smallest contribution from the 3-dimensional
screened Coulomb t-matrix 
  $ \left\langle \Phi_0 | (1+P)t_c^RP | {\Phi } \right\rangle $  
brings the theory to the pd data for the QFS geometry at all S-values and
 for the SST  configuration at S-values close to the space-star condition.

\section{Summary and conclusions}
\label{sumary}

We extended our  approach  to include the pp Coulomb
force into the momentum space 3N Faddeev calculations presented 
in Ref.~\cite{elascoul} for elastic scattering to the pd breakup. 
It is based on a standard
 formulation for short range forces and relies on the screening  of the
long-range Coulomb interaction. In order to avoid  all uncertainties
connected with the application of the partial wave expansion, 
unsuitable when working
with long-range forces, we apply directly the 3-dimensional pp screened
Coulomb t-matrix.

Using a simple dynamical model for the nuclear part of the interaction
we demonstrated feasibility of that approach for the 
treatment of the pd breakup. 
 We demonstrated that contrary to the pd elastic scattering, where 
 the resulting amplitudes do not require renormalisation, 
 it is unavoidable to perform renormalisation of the pp half-shell t-matrices 
in order to get the physical  breakup amplitude. Namely that 
amplitude has two contributions, one driven by the interaction in the
pp subsystem and second in the np subsystem. Only the first part 
requires renormalisation.

We have shown that converged results for 
breakup can be achieved with finite screening radia. 

We calculated contributions of different terms to the breakup cross section 
 in QFS and SST
configurations.  
 The interference
between different contributions leads to an interference pattern, which is  
different for QFS and SST configurations. 
 In our restricted dynamical model the pp Coulomb interaction brings  
 the nd breakup cross sections close to the pd data for the QFS
configuration. Also for the SST geometry in the vicinity of the space-star 
condition the pd theory is close to the pd data. However,  further away on 
 the S-curve the theory lies above the data. 

In future studies the perturbative approximation for the 3-dimensional
screened Coulomb t-matrix will be avoided and higher partial wave
components of the nucleon-nucleon interaction will be included.

\section*{Acknowledgments}
This work was supported by the Polish 2008-2011 science funds as a
 research project No. N N202 077435.
It was also partially supported by the Helmholtz
Association through funds provided to the virtual institute ``Spin
and strong QCD''(VH-VI-231)  and by
  the European Community-Research Infrastructure
Integrating Activity
``Study of Strongly Interacting Matter'' (acronym HadronPhysics2,
Grant Agreement n. 227431)
under the Seventh Framework Programme of EU. 
 The numerical
calculations have been performed on the
 supercomputer cluster of the JSC, J\"ulich, Germany.

\end{document}